\documentclass{article}   
\usepackage{epsf}  
 
\begin{document}

\centerline{\bf Quantum Mechanics of  a Simulated Trihydrogen Dication}
\vskip .5in
\centerline{ML Glasser}\vskip .2in

\centerline{ Departamento de F\'isica
Te\'orica, At\'omica y \'Optica, }
\centerline{ Universidad
de Valladolid\\
47071 Valladolid, Spain}
\centerline{ Department of Physics, Clarkson University, }
\centerline{Potsdam, NY 13699-5820 (USA)}

\begin{abstract}
 The Schroedinger equation is solved exactly within the Born-Oppenheimer approximation for a simulacrum of the $H_3^{++}$-ion. The ion is assumed to form an isosceles triangle and the ground state energy is obtained over its geometrical parameter space. No multi-center molecular integrations are required.   We indicate how the approximation to the actual molecule can be improved systematically. \end{abstract}

\noindent
pacs{ 31.15.Ar; 02.70.Wz; 31.50.Bc


\section{Introduction}
 The Trihydrogen cation $H_3^+$ was identified, by mass spectroscopy, in 1911 by J.J. Thomson[1]. Twelve years later Hogness and Lunn[2] found that it could be produced by the proton exchange $H_2^++H_2\rightarrow H_3^{+}+H$ and would readily lose an electron. Subsequently it was found that  $H_3^+$ is present in interstellar clouds and is among the most abundant molecular species in the universe. This led to the question of the stability of the dication $H_2^{++}$, which has remained somewhat controversial to the present, though the consensus is that it is unstable. 
 
 The first quantum treatment was by Gordadse[3, 4] in (1935), who assumed the protons were  fixed, equally spaced along a straight line or formed an equilateral triangle. He used the variational method, as have all subsequent studies,  based on a  one-parameter trial function built from the hydrogen 1s-state.  In spite of the  simplicity of the trial functions, he found several of the multi-center integrations intractable, requiring not too well controlled approximations, and concluded that neither configuration was stable. About the same time C.A. Coulson carried out a LCMO study of $H_3^{++}$ [5]. He used only a single  molecular orbital and his energy values lie somewhat higher than those in [4].  The difficulty of these multi-center integrations have continued to dog such calculations  and may be what prompted Eyring's comment that $H_3^+$ is ``the scandal of modern chemistry" [6].  The first extensive study of $H_3^{++}$ using electronic digital computers was by  H. Conroy in 1964 [7,8,9,10], who devised an insightful set of variational wave functions and used Monte-Carlo algorithms for the integrations, remarking that ``these integrations presented grave obstacles". This and subsequent calculations [11,12,13,14,15], mostly confined to the linear and equilateral triangle configurations, have upheld Gordadse's conclusion that the ion is unstable.  An attempt to produce the dication experimentally [16, 17] was unsuccessful.
 
    The purpose of this note is to suggest that it is possible to avoid the variational method and consequent multi-center integrations entirely and that  the Schroedinger equation  can be  solved exactly for a sequence of Hamiltonians that converge to the correct one.  This note is intended  as a proof of principle, and deals only with the first Hamiltonian in the sequence; it is equivalent to keeping  only  the hydrogen 1s state and is simple enough that most of the calculations can be done ``by hand".  We also assume that the ion forms an isosceles triangle and the ground-state energy (ignoring hyperfine effects) is examined as a function of a side and adjacent angle. Even so, the results lie reasonably close to the most recent values.

\section{Hydrogen atom}
Let the one-electron Hamiltonian of an atom ( or indeed, any system) be 
$$H=p^2+V(\vec{r})\eqno(1.1)$$
and have bound-state eigenfunctions and energy levels $\{\phi_a,\; E_a\}$. Then, by completeness,
$$V(\vec{r})=\sum_av_a(\vec{r})<\phi_a|,\qquad v_a(\vec{r})=V(\vec{r})|\phi_a>.\eqno(1.2)$$
Thus, if we ignore the continuum states, which will be of no interest in the sequel, then we have a sequence of Hamiltonians
$$H_n=p^2+\sum_{m=0}^nv_m<\phi_m|\eqno(1.3)$$
and it is easily checked that the Schroedinger equation (we adopt units: $\hbar=2m=e^2/2=1$)
$$\{-\nabla^2-E\}\psi(\vec{r})=-\sum_{m=0}^nv_m(\vec{r})\lambda_m$$
$$\lambda_m=\int d\vec{s} \phi_m^*(\vec{s})\psi(\vec{s})\eqno(1.4)$$
has precisely the first $n$ of the eigenstates of $H$. By transforming to momentum space and writing $E=-\epsilon<0$ (since we are only concerned with bound states) (4) becomes the integral equation
$$\hat{\psi}(\vec{k})=\sum_{m=0}^n\frac{\hat{v}_m(\vec{k})\lambda_m}{\epsilon+k^2}.\eqno(1.5)$$
Next, by the Parseval relation for the Fourier transform [2] we have the consistency condition
$$\lambda_q=\frac{1}{(2\pi)^3}\int \hat{\phi}_q^*(\vec{k})\hat{\psi}(\vec{k}) d\vec{k}=\sum_m A_{qm}\lambda_m$$
$$A_{qm}=\frac{1}{(2\pi)^3}\int \frac{\hat{v}_m(\vec{k})\hat{\phi}_q^*(\vec{k})}{k^2+\epsilon}d\vec{k}.\eqno(1.6)$$
That is to say, the energy levels $\epsilon$ and corresponding  $\lambda$'s are determined by the matrix equations
$$({\bf A}-{\bf I})\Lambda=0,\qquad Det|{\bf A}-{\bf I}|=0\eqno(1.7)$$
where $ \Lambda$ is the column vector $(\lambda_0,\cdots,\lambda_n)^T$.  In the next section we illustrate this by working out  $H_0$ and $H_1$ for hydrogen.

   It should be pointed out that in (1.2) the $\phi_a$ may be  any complete set of functions, not just the eigenfunctions of $H$, though the low order approximations are unlikely to be as accurate.

\subsection{Hydrogen atom}

The lowest two bound-state wave functions for Hydrogen are[3]
$$\phi_0(r)=\pi^{-1/2}e^{-r}, \quad \epsilon_0=1, \mbox{ and  }  \phi_1(r)=(32\pi)^{-1/2}(2-r)e^{-r/2},\quad \epsilon_1=\frac{1}{4}.\eqno(2.1)$$
Hence,
$$\hat{\phi}_0(k)=\frac{8\sqrt{\pi}}{(k^2+1)^2},\quad \hat{\phi}_1(k)=\frac{32\sqrt{2\pi}(4k^2-1)}{(4k^2+1)^3}$$
$$\hat{v}_0(k)=\frac{8\sqrt{\pi}}{(k^2+1)},\quad \hat{v}_1(k)=\frac{8\sqrt{2\pi}(4k^2-1)}{(4k^2+1)^2}\eqno(2.2)$$
and the matrix elements of ${\bf A}$ are ($\epsilon=x^2$)
$$A_{00}=2\frac{3+x}{(1+x)^3}$$
$$A_{11}=2\frac{8x^3+20x^2+6x+7}{(2x+1)^5}\eqno(2.3)$$
$$A_{01}=\frac{32\sqrt{2}}{27}\frac{2x^2+5x-7}{(1+x)(1+2x)^3}$$
$$A_{10}=\frac{8\sqrt{2}}{27}\frac{4x^2+12x-7}{(1+x)^2(1+2x)^2}.$$
 It is not difficult to check that the determinant in (1.7) has the form $(x-1)(2x-1)P(x)/Q(x)$ where $P$ and $Q$ are polynomials with positive coefficients, so its sole positive real roots are $x=1/2$ and $x=1$. The  first equation  (1.7) has the solution $\lambda_0=1$, $\lambda_1=A_{10}/(1-A_{00})$
and it is straightforward to check that inverting (1.5) with these values of $x$  reproduces (2.1).

\section{Triangular Molecule}

Consider the three-proton system, in the Born-Oppenheimer approximation, where one lies at the origin and two lie in the $x-z$-plane at positions
$$\vec{R}_{\pm}=R(\cos\alpha,0,\pm \sin\alpha).\eqno(3.1)$$
The Schroedinger equation for an electron subject to this configuration is
$$-(\nabla^2+E)\psi(\vec{r})=[V(\vec{r})+V(\vec{r}-\vec{R}_+)+V(\vec{r}-\vec{R}_-)]\psi(\vec{r}).\eqno(3.2)$$
Setting $E=-\epsilon$, (3.2) has the immediate solution in momentum space
$$\hat{\psi}(\vec{k})=\frac{\hat{v}_0(\vec{k})}{k^2+\epsilon}[\lambda_0+e^{i\vec{k}\cdot\vec{R}_+}\lambda_++e^{i\vec{k}\cdot\vec{R}_-}\lambda_-]\eqno(3.3)$$
with
$$\lambda_0=\frac{1}{(2\pi)^3}\int d\vec{k} \hat{\phi}_0(\vec{k})^*\hat{\psi}(\vec{k})$$
$$\lambda_{\pm}=\frac{1}{(2\pi)^3}\int d\vec{k}\hat{\phi}_0(\vec{k})^*e^{-i\vec{k}\cdot\vec{R}_{\pm}}\hat{\psi}(\vec{k}).\eqno(3.4)$$

Therefore, by defining the four integrals
$$I_0=\frac{1}{(2\pi)^3}\int d\vec{k}\frac{\hat{\phi}_0^*(\vec{k})\hat{v}_0(\vec{k})}{k^2+\epsilon}$$
$$I_{\pm}=\frac{1}{(2\pi)^3}\int d\vec{k}\frac{\hat{\phi}_0^*(\vec{k})\hat{v}_0(\vec{k})}{k^2+\epsilon}e^{i\vec{k}\cdot\vec{R}_{\pm}}\eqno(3.5)$$
$$I_1=\frac{1}{(2\pi)^3}\int d\vec{k}\frac{\hat{\phi}_0^*(\vec{k})\hat{v}_0(\vec{k})}{k^2+\epsilon}e^{-i\vec{k}\cdot(\vec{R}_+-\vec{R}_-)},$$
we have the three consistency equations
$$\lambda_0=I_0\lambda_0+I_+\lambda_++I_-\lambda_-$$
$$\lambda_+=I_+^*\lambda_0+I_0\lambda_++I_1\lambda_-\eqno(3.6)$$
$$\lambda_-=I_-\lambda_0+I_1^*\lambda_++I_0\lambda_-.$$

 The four integrals (3.5) are real, $I_+=I_-$ and is independent of $\alpha$. From (3.6) we see that the ground state energy $\epsilon$ is fixed by the determinantal equation ($\epsilon=x^2)$
$$F[R,\epsilon]\equiv
(I_0-I_1-1)[(I_0-1)^2+(I_0-1)I_1-2I_+^2]=0.\eqno(3.7)$$
The integrals (3.5) are elementary:
$$I_0=\frac{2(3+x)}{(1+x)^3}$$
$$I_{\pm}=f(R,x)=\frac{16}{(x^2-1)^2}\left\{\frac{e^{-R}-e^{-Rx}}{R(x^2-1)}+\frac{1}{8}e^{-R}[x^2-5+R(x^2-1)]\right\}\eqno(3.8)$$
$$I_1=f(2R\sin\alpha,x).$$

The spatial wave function is obtained through the Fourier inversion of (3.3) by which we find in spherical coordinates
$$\psi(r,\theta,\phi)=$$
$$N\left\{\frac{e^{-r}-e^{-xr}}{r}+\left(\frac{1-I_0}{2I_+}\right)\left[\frac{e^{-\rho_+}-e^{-x\rho_+}}{\rho_+}+\frac{e^{-\rho_-}-e^{-x\rho_-}}{\rho_-}\right]\right\},\eqno(3.9)$$
where $N$ is a normalization factor and
$$\rho_{\pm}=\sqrt{r^2+r^2-2rR(\sin\theta\cos\varphi\cos\alpha\pm \cos\theta\sin\alpha)}.\eqno(3.10)$$
The wave function in the equilateral triangle configuration for $r=R=1.6$,  is shown as a function of $\theta$ and $\varphi$ in Fig.1.

 \begin{figure}
\begin{center}
\epsfbox{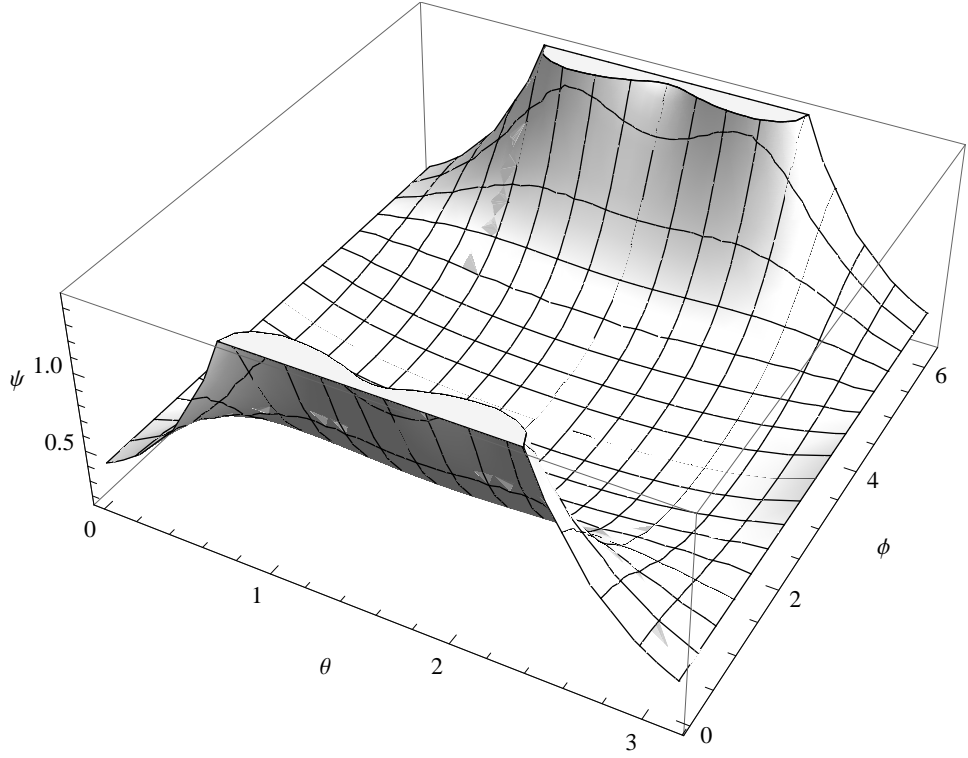}
\end{center}
\caption{$\psi(1.6,\theta,\phi)$ for $0<\theta<\pi$, $0<\phi<2\pi$.}
\end{figure}

\section{Results and discussion}

The ground state energy $E=-\epsilon=-x^2$ is given by the largest positive root $x$ of (3.7) which is that of the second factor. This is most easily determined graphically and the results for four cases are given below.

\subsection{Linear configuration: $\alpha=\pi/2$.}

$$\begin{array}{ccccc}

  R&x&&R&x\\
  
  0&2.1349367&& 1.2&1.8036179\\
  
  0.1& 2.1307780&&1.4&1.7347596\\
  
  0.2& 2.1187612&&  1.6& 1.6694538 \\
  
 0 .3&2.0999631 &&1.8& 1.6082098 \\
 
 0.4& 2.0756736&&2.0 &1.5512046\\

  0.5& 2.0471617&&  2.2& 1.4984244\\
  
   0.6& 2.0155626&&  2.4& 1.4497481 \\
   
    0.7& 1.9818363 && 2.6& 1.4049973\\
    
    0.8& 1.9467667 && 2.8& 1.3639650 \\
    
    1.0& 1.8749557&& 3.0& 1.3264326
    
    \end{array}$$

The ground-state energy vs $R$ for $\alpha=\pi/2$ is shown in Fig.2 and the total molecular energy in Fig.3; the ion is unstable.

 \begin{figure}
\begin{center}
\epsfbox{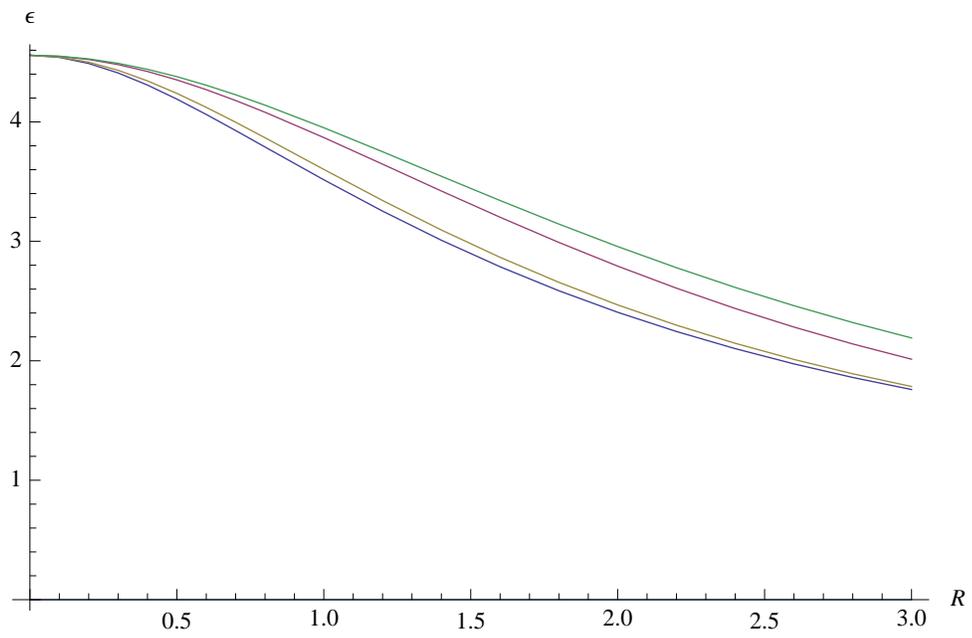}
\end{center}
\caption{Ground-State energy $\epsilon$ vs $R$. From top to bottom at $R=1$: $\alpha=\pi/8$, $\alpha=\pi/6$, $\alpha=\pi/3$, $\alpha=\pi/2$}
\end{figure}

 \begin{figure}
\begin{center}
\epsfbox{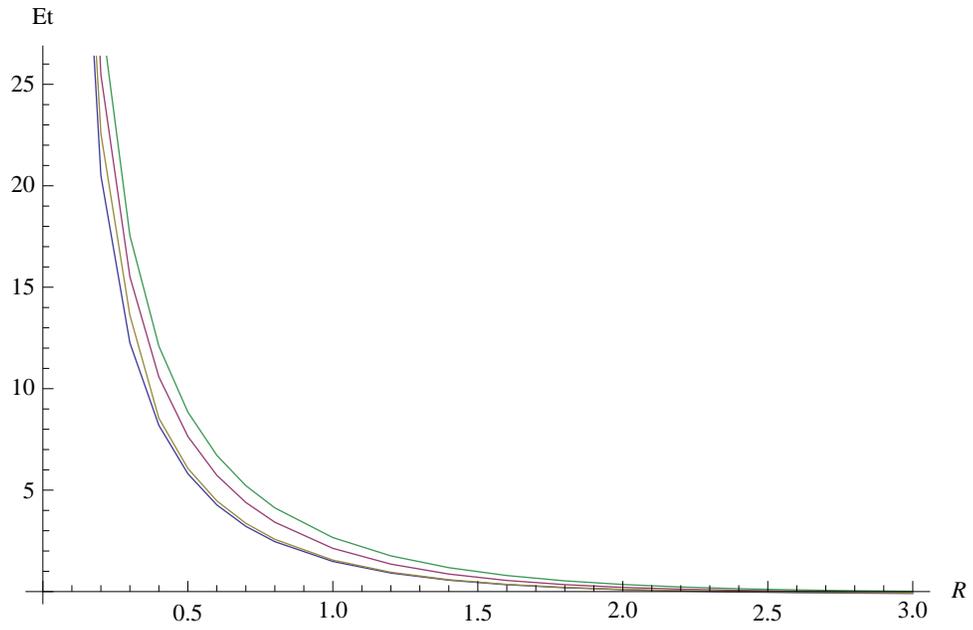}
\end{center}
\caption{ Total energy vs. $R$: Same order as in Fig.1.}
\end{figure}

 \subsection{ Equilateral triangle $\alpha=\pi/6$}
 
 $$\begin{array}{ccccc}
R&x&&R&x\\

0.0& 2.1349367 &&  1.2 &1.9095172 \\

0.1& 2.1328424&& 1.4 &1.8497556 \\

0 .2& 2.1266444 &&1.6 &1.7893258\\

0 .3& 2.1165658&& 1.8 &1.7294701 \\
 
 0 .4& 2.1029191 && 2.0& 1.6711139 \\

 0 .5& 2.0860668 && 2.2&  1.6149246 \\
  
 0 .6&  2.0663935&& 2.4 &1.5613632\\
  
 0  .7& 2.0442855 &&  2.6 &1.5107287 \\
   
  0 .8& 2.0201173&& 2.8 & 1.4631940 \\
   
    1.0 &1.9669892 &&3.0 &1.4188350
    
\end{array}$$  
 
 The ground-state  and total energies are shown as functions of $R$ in Fig.2 and Fig.3 For this geometry with $R=1.68$ the exact ground-state energy $x=1.95426$ has been proposed[15]. Our value at this spacing is $x=1.76526$ a difference of just under $10\%.$ In Fig.4. we show our result for the total energy $Et$ compared to a recent study of the equilateral case by  Medel-Cobaxin et al.[18] \begin{figure}
 
\begin{center}
\epsfbox{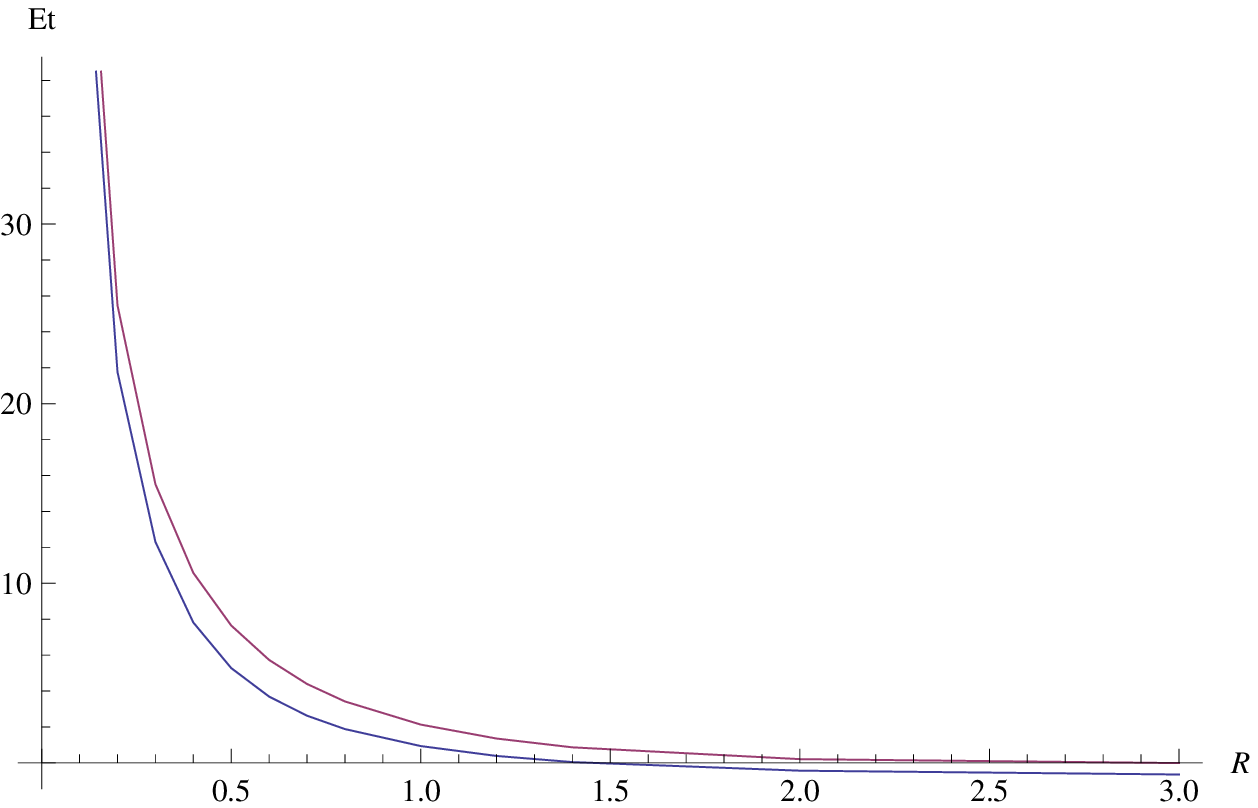}
\end{center}
\caption{Total energy vs $R$:Comparison with Ref.[18]. Upper curve--this work.}
\end{figure}

 \subsection{ Isoceles cases: $\alpha=\pi/3,\, \pi/8$}
 
 For completeness we show the total molecular energy for an  obtuse ($\alpha=\pi/3$ ) and an acute ($\alpha=\pi/8$) triangular configuration in Figs.2-3. Again, in neither case is the ion stable.\vskip .1in
 
 \noindent{\bf $\alpha=\pi/3$}
 
   $$\begin{array}{ccccc} 
  R&x&&R&x\\ 
   
   0.0& 2.1349367 &&  1.2 &1.8281244 \\
   
   0.1& 2.1314613&& 1.4& 1.7592914 \\
   
   0 .2& 2.1213259 &&1.6 &1.6929713 \\
    
    0.3 &2.1052384&& 1.8 &1.6300407\\
  
  0 .4 &2.0840921 && 2 .0&1.5709739\\
   
   0.5& 2.0588167 &&  2.2 &1.5159789 \\
   
  0 .6& 2.0302939&& 2.4 &1.4650897 \\
  
 0  .7 &1.9993156 && 2.6 &1.4182289\\
   
  0 .8 &1.9665685 && 2.8 &1.3752500\\
   
   1. 0&1.8979968 &&3.0 &1.3359649
   
 \end{array}$$

\vskip .2in
\noindent{\bf $\alpha=\pi/8$}

  $$\begin{array}{ccccc }
  R&x&&R&x\\

  0.0& 2.1349367&&1.2& 1.9365603\\
  
   0.1& 2.1331309&& 1.4& 1.8828383\\
   
   0.2& 2.1277809&& 1.6&  1.8280552\\
   
   0 .3&  2.1190636&&1.8& 1.7733225\\ 
   
   0.4& 2.1072292&& 2.0& 1.7194870\\ 
   
    0.5& 2.0925704&& 2.2&  1.6671775\\
    
     0.6&  2.0754001&&   2.4& 1.6168452\\ 
   
   0.7& 2.0560355&& 2.6& 1.5688001\\
   
    0.8& 2.0347866&& 2.8& 1.5232405\\
    
     1.0& 1.9877994&& 3.0& 1.4802771
   
    \end{array}$$

In conclusion, we have given the {\it exact} solution of the Schroedinger equation, within the Born-Oppenheimer approximation,  for a model three-center molecule closely resembling $H_3^{++}$. For the equilateral configuration, where an exact ground-state energy  at $R=1.68$ has been proposed[15] the value calculated here agrees to better than 10\%.. Furthermore, for our model:

         \begin{itemize}
         
    \item No multi-center molecular integrals are required.
         
     \item The approximation can be systematically improved.
     
     \item The  corresponding Dirac equation can be solved exactly [20].
     
     \item Electric and magnetic fields can be included requiring only  the solution of a first or second order ODE.[21] 

     \item It may be feasible to treat the Kohn-Sham equations on the same basis, in which case correlation effects can be included. 
 
  \end{itemize}

\noindent
{\bf Acknowledgements}\vskip .1in

 I thank the Universidad de Valladolid, Valladolid, Spain and the DHIP, San Sebastian, Spain, where  this work was carried out, for their hospitality. I am also grateful to Prof. Carlos Balbas for comments and references.

\end{document}